\documentclass[12pt,preprint]{aastex}
\usepackage{natbib}
\usepackage{graphicx}
\begin{document}
\title{Near-infrared spectroscopy of the Blue Compact Dwarf galaxy Markarian 59
}
\author{Yuri I. Izotov}
\affil{Main Astronomical Observatory, Ukrainian National Academy of Sciences,
27 Zabolotnoho str., Kyiv 03680, Ukraine}
\email{izotov@mao.kiev.ua}
\author{Trinh X. Thuan}
\affil{Astronomy Department, University of Virginia, P.O. Box 400325, 
Charlottesville, VA 22904-4325}
\email{txt@virginia.edu}
\and
\author{John C. Wilson}
\affil{Astronomy Department, University of Virginia, P.O. Box 400325, 
Charlottesville, VA 22904-4325}
\email{jcw6z@virginia.edu}

\begin{abstract}
We present near-infrared (NIR) spectroscopic observations 
of the blue compact dwarf 
(BCD) galaxy Mrk 59,   
obtained with the TripleSpec spectrograph mounted on the 3.5m APO 
telescope. The NIR spectrum of Mrk 59, which covers 
the 0.90 $\mu$m -- 2.40 $\mu$m
wavelength range, shows atomic hydrogen, 
molecular hydrogen, helium, sulfur and iron emission
lines. The NIR data have been supplemented by a SDSS 
optical spectrum. We found extinction in the BCD 
to be low [$A(V)$=0.24 mag] 
and to be the same in both 
the optical and NIR ranges. The NIR light does not reveal hidden 
star formation. The  H$_2$ emission
comes from dense clumps and 
the H$_2$ vibrational emission line intensities can be accounted for by
photon excitation. No shock excitation is needed. 
 A CLOUDY photoinization model of Mrk 59 
reproduces well the
observed optical and NIR emission line fluxes. There is no need
to invoke sources of ionization other than stellar radiation.
The [Fe {\sc ii}] 1.257 and 1.643 $\mu$m emission lines, often used 
as supernova shock indicators in low-excitation high-metallicity starburst 
galaxies, 
cannot play such a role in high-excitation low-metallicity 
H {\sc ii} regions such as Mrk 59.

\end{abstract}


\section{INTRODUCTION}

The Blue Compact Dwarf (BCD) galaxy Markarian 59 (Mrk 59) $\equiv$ I Zw 49 
belongs to the 
class of cometary BCDs defined by \citet{LT85} as 
characterized by an intense starburst at the end of an elongated 
low surface brightness stellar 
body. In the case of Mrk 59, the elongated body is the galaxy NGC 4861. 
\citet{A66} describes NGC 4861 in his Atlas of Peculiar Galaxies as ``an 
object with irregular clumps, resolved into knots with a very bright knot 
(diameter = 1 kpc) at the southeastern end''.  
The knots are in fact a chain of 
H {\sc ii} regions, most likely resulting from propagating star formation along 
the galaxy's elongated body and ending with the high-surface-brightness 
supergiant H {\sc ii} region at the southeastern end \citep{N00}. 
Fabry-P\'erot maps of NGC 4861 by \citet{T87} show that Mrk 59 is 
$\sim$ 50 times brighter in H$\alpha$ than any other feature in NGC 4861. The 
H$\alpha$ velocity field of NGC 4861 shows a northeast-southwest velocity gradient 
which can be interpreted as the rotation of an inclined disk, with the bright 
end approaching and the faint end receding.   
  
 \citet{DS86} first detected broad
Wolf-Rayet stellar features in Mrk 59 indicating the presence of late nitrogen
and early carbon Wolf-Rayet stars. \citet{G00}, 
using the spectroscopic observations of \citet{I97} 
found that several dozens of Wolf-Rayet stars are present. 
\citet{N00} used deep ground-based spectrophotometric observations  
of the supergiant H {\sc ii} region to 
derive an oxygen abundance (log O/H)$_i$ = $-$4.011 $\pm$ 0.003 
[or $Z_\odot$/5, using log (O/H)$_\odot$ = $-$3.3] for the ionized 
gas, typical of BCDs. 
On the other hand, \citet{T02b} have found
from {\sl FUSE} far-UV absorption spectra that the oxygen abundance of the neutral
gas in Mrk 59 is about a factor of ten lower, (log O/H)$_n$ = --5.0 $\pm$ 0.3. They also
did not find any evidence for the presence of diffuse molecular hydrogen in the
BCD.
\citet{W96} and \citet{T02a} have used the VLA to map NGC 4861 and 
Mrk 59 in the 21 cm line. With a beam size of 15\arcsec\ (780 pc), the latter 
authors found a very clumpy interstellar medium (ISM) 
 with H {\sc i} column densities ranging from several 10$^{19}$ to a few 
10$^{21}$ cm$^{-2}$. 
Despite irregularities, the H {\sc i} velocity field of NGC 4861 resembles 
that of a rotating disk seen almost edge-on.
\citet{T02a} obtained 
a H {\sc i} heliocentric velocity of 833 km s$^{-1}$ for NGC 4861. Then, adopting 
a Hubble constant of 73 km s$^{-1}$ Mpc$^{-1}$ and  
correcting for the Virgocentric infall motion,  
the distance of Mrk 59 is 10.7 Mpc \citep{T02a}. 
Since the apparent magnitude of Mrk 59 is $m_B$ = 12.64 \citep{N00}, then 
its absolute magnitude $M_B$ is equal to --17.51.

In the near-infrared (NIR), NGC 4861 and Mrk 59 have been studied 
spectroscopically by \citet{C96} and \citet{VR97}, respectively. 
\citet{C96} \citep[see also ][]{C97} have obtained spectra of NGC 4861 in the 
wavelength range
1.236 - 1.340 $\mu$m ($J$ spectrum) and 2.113 - 2.250 $\mu$m
($K$ spectrum). They detected the 
[Fe {\sc ii}] 1.26 $\mu$m, Pa$\beta$ 1.28 $\mu$m, Br$\gamma$ 2.17 $\mu$m
and H$_2$ 2.12 $\mu$m lines.
\citet{VR97} have obtained a $K$ spectrum of Mrk 59 covering 
the 2.0 -- 2.4 $\mu$m wavelength range. Those authors detected only 
three emission lines: the Br$\gamma$ 2.17 $\mu$m hydrogen line, the 
He {\sc i} 2.06 $\mu$m line and the H$_2$ 2.12 $\mu$m line, with the last 
line being barely detected. With advances in NIR detector sensitivities, 
and with the commisionning in 2008 of the new TripleSpec spectrograph at the 
Apache Point Observatory (APO) 3.5-meter telescope\footnote{The Apache Point 
Observatory 
3.5-meter telescope is owned and operated by the Astrophysical Research 
Consortium.}, which allows simultaneous coverage of the $JHK$ NIR bands,
we decided to obtain a new NIR spectrum of Mrk 59 with considerably more 
signal-to-noise and wavelength coverage. These improved observations will  
allow us to study in more detail the physical conditions in Mrk 59:  
its extinction in the NIR and the excitation mechanisms
of line emission of ionized species in the H {\sc ii} region.
We can also check for the presence of molecular hydrogen emission. 

We describe the 3.5m APO observations in section 2. In section 3, 
we discuss the extinction and star formation in Mrk 59, both in the optical and NIR ranges, 
the excitation mechanisms for 
molecular hydrogen and iron emission in the NIR range, and CLOUDY modeling 
of the H {\sc ii} region.
We summarize our findings in section 4.

\section{OBSERVATIONS}

A NIR spectrum of Mrk 59 was obtained with the 3.5 m APO telescope on the 
night of 2008 March 22. The observations were made with the TripleSpec 
spectrograph by the instrument
commissioning team. TripleSpec \citep{W04} is a cross-dispersed NIR spectrograph that 
provides simultaneous continuous wavelength coverage from 0.95 to 2.46 $\mu$m in
five spectral orders during a single exposure. 
A 1\farcs1$\times$43\arcsec\ slit was used, resulting in a resolving power
of 3500. 

The A0/A1IV standard star HIP 73613 was observed for flux
calibration. Spectra of Ar comparison arcs were obtained on the same date
for wavelength calibration. Since both Mrk 59 and HIP 73613 
are much smaller than the length of the slit, the nod-on-slit 
technique was used to acquire the sky spectrum. Both objects were observed
by nodding between two positions A and B along the slit. 
Mrk 59 had a total integration 
time of 20 minutes. During that time the galaxy was observed in four 
five-minute 
integrations, alternating in slit position following the sequence ABBA.

The data reduction was performed using IRAF\footnote{IRAF is 
the Image Reduction and Analysis Facility distributed by the 
National Optical Astronomy Observatory, which is operated by the 
Association of Universities for Research in Astronomy (AURA) under 
cooperative agreement with the National Science Foundation (NSF).}
and include the following steps.
We first used the summed frame 
of the standard star and the IRAF
routines IDENTIFY, REIDENTIFY, FITCOORD and TRANSFORM to correct 
for the distortion of each frame. This distortion is an unavoidable 
consequence of cross-dispersed grating spectrographs. 
This was done for all frames of the
standard star, Mrk 59 and the comparison argon lamp. We then extracted from
each of the distortion-corrected 
frames the five two-dimensional spectra corresponding
to the five spectral orders. For each order, we use the two-dimensional 
distortion-corrected spectrum of the comparison argon lamp and the 
IRAF routines 
IDENTIFY, REIDENTIFY, FITCOORD, TRANSFORM to perform wavelength calibration 
and correct for the tilt of the slit for each frame. We checked the
accuracy of the two-dimensional transformation by examining the strong lines
in the A and B spectra of Mrk 59. We found
that the wavelengths of these 
do not differ by more 
$\sim$ 0.1 -- 0.2\AA\ in the A and B spectra. 
Then, for each order, the two A and two B 
distortion-corrected, tilt-corrected and wavelength-calibrated frames, each with an 
exposure time of 5 mn, were summed separately. The B+B frame was subtracted 
from the A+A frame to remove the background. Then, the -(B+B) spectrum 
was shifted along the spatial axis to match the (A+A) spectrum and 
subtracted from it, resulting in the two-dimensional background subtracted 
spectrum (A+A)+(B+B) of Mrk 59 with
a total exposure time of 1200 s. The standard star observations were
processed similarly, resulting in a spectrum with a total exposure time of
300 s. Finally, we use the routine CRMEDIAN to remove cosmic ray hits in each
spectrum.

\begin{figure*}
\figurenum{1}
\hbox{\includegraphics[angle=0,width=0.9\linewidth]{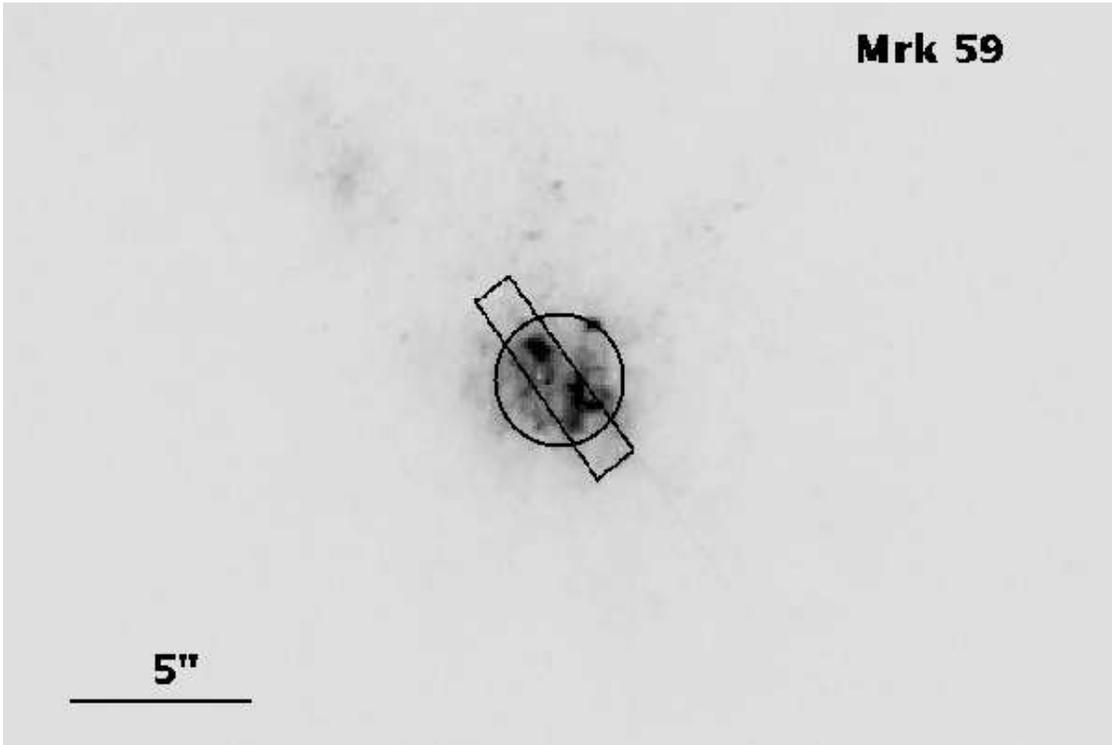} 
}
\figcaption{The
extraction aperture of 1\farcs1$\times$6\arcsec\ for the one-dimensional
NIR spectrum is shown by a rectangle. The  3\arcsec\ diameter circular 
aperture of the SDSS optical spectrum is also shown. Both NIR and optical 
spectra are overlayed on an archival {\sl HST}/WFPC2 image of Mrk 59. North is up and East is to the left.
\label{fig1}}
\end{figure*}

A one-dimensional non-flux-calibrated spectrum of Mrk 59 was extracted from 
the two-dimensional frame using the APALL IRAF routine, within a 
1\farcs1$\times$6\arcsec\ rectangular aperture so as to include the
brightest star-forming regions. The extraction aperture is shown 
in Fig. \ref{fig1}, overlaid on an archival {\sl HST}/WFPC2 image of Mrk 59.
We will hereafter refer to it as the NIR aperture. 

We simultaneously correct for telluric absorption and flux-calibrate 
the spectrum of Mrk 59 by dividing the 
non flux-calibrated spectrum of Mrk 59 by that of the standard star
HIP 73613 and by multiplying the result by the synthetic flux-calibrated 
spectrum of that same standard star.
Since there does not exist a published absolute spectral energy distribution
of the standard star HIP 73613, we have used 
the synthetic absolute spectral energy 
distribution of the star Vega ($\alpha$ Lyrae) which has a 
similar spectral type (A0V), and scaled it to the brightness of HIP 73613.

\section{RESULTS AND DISCUSSION}

\begin{figure*}
\figurenum{2}
\hbox{\includegraphics[angle=0,width=0.9\linewidth]{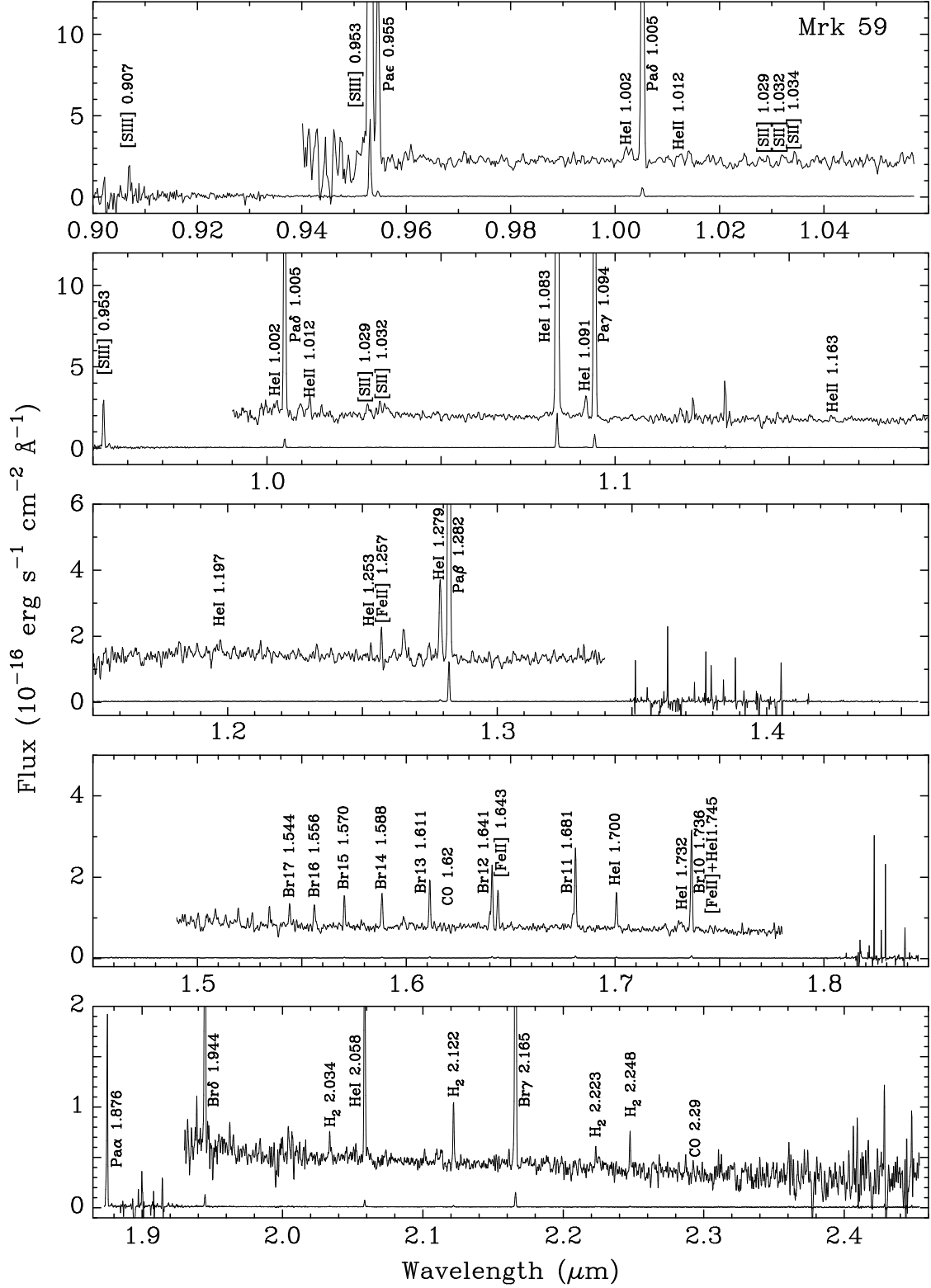} 
}
\figcaption{The 3.5m APO/TripleSpec NIR spectrum of Mrk 59 in five orders. 
In each panel, the noisy regions of the upper spectrum are omitted.  
They are caused by insufficient sensitivity or strong telluric
absorption. The flux scale on the y-axis corresponds 
to the upper spectrum. The lower spectrum is downscaled by a factor of 50
as compared to the upper spectrum. It is shown for the whole wavelength
range in each order.
\label{fig2}}
\end{figure*}

The resulting rest-frame NIR spectra of Mrk 59, 
corrected for telluric absorption, are shown in Fig. \ref{fig2}.
Each of the five orders is shown in a separate panel,
with the identified lines labeled. Each panel shows two spectra.
The upper spectrum is the blown-up version, corresponding to the 
flux scale given on the y-axis. It excludes the noisy parts. 
The lower spectrum is the downscaled version of the upper one by a factor 
of 50. It shows the whole wavelength range of the spectrum in  
each order. 

\begin{figure*}
\figurenum{3}
\hbox{\includegraphics[angle=-90,width=1.0\linewidth]{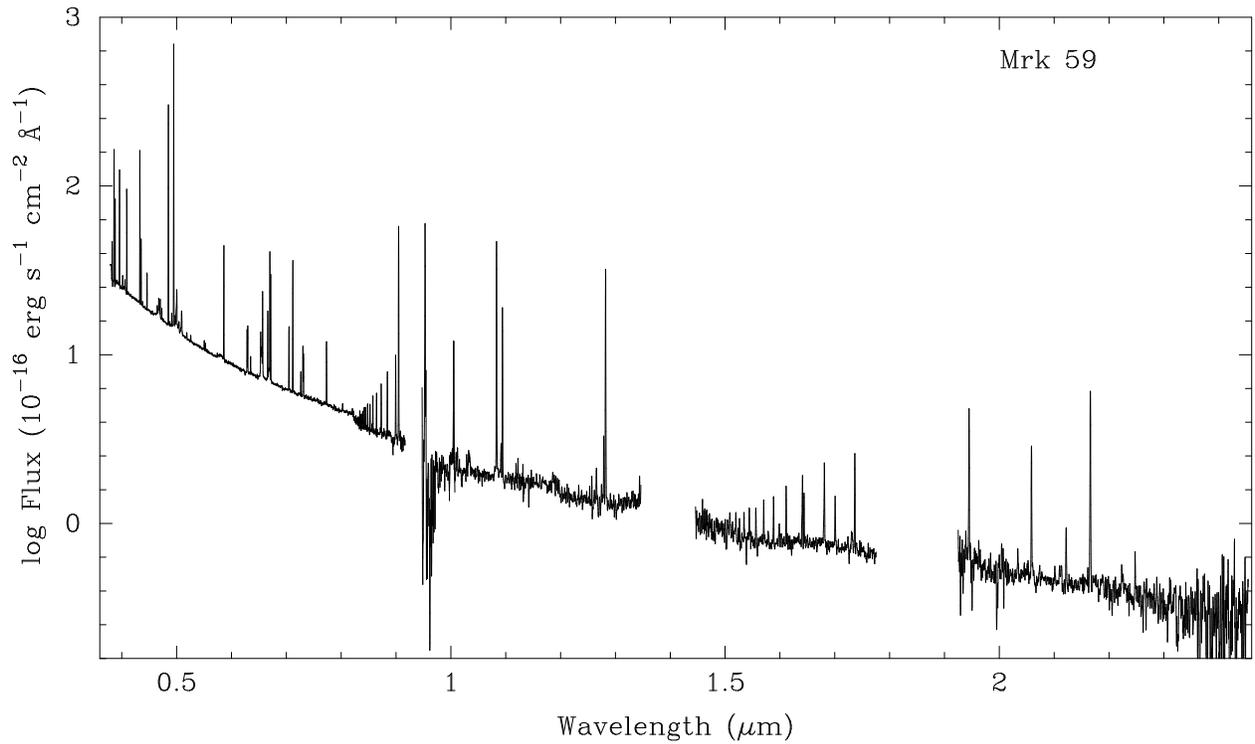}} 
\figcaption{Spectral energy distribution of Mrk 59 in the wavelength
range 0.39 $\mu$m -- 2.4 $\mu$m. The optical spectrum in the wavelength range 
0.39 $\mu$m -- 0.93 $\mu$m is from the SDSS. The other spectra have been obtained with
the 3.5m APO telescope and the TripleSpec spectrograph. \label{fig3}}
\end{figure*}

To have a more complete physical picture of the extinction and 
the ionization mechanisms in Mrk 59, we have supplemented the 
NIR observations with 
optical ones. We have used the optical spectrum of Mrk 59 in the Sloan Digital Sky Survey (SDSS) archives obtained within a circular aperture
of 3\arcsec\ diameter, which 
covers the wavelength range 0.38 -- 0.93 $\mu$m.
 The optical and NIR apertures are centered on the same spatial location 
in Mrk 59 and include the same brightest star-forming 
regions (see Fig. \ref{fig1}), so the optical and NIR spectra 
are directly comparable. They overlap in
the wavelength range of 0.90 -- 0.93 $\mu$m, allowing us to
match one to the other. Specifically, we have adjusted 
the optical and NIR spectra by scaling the latter by a factor 1.03, 
so that the flux of the [S {\sc iii}] 0.907 $\mu$m emission line is 
the same in both spectra. This procedure works well  
despite the fact that this line is located in a noisy part of the NIR spectrum
and its flux is derived with an error of 10\% (the flux error of 
the same line in the optical spectrum is smaller, of the order of 3\%).
The combined redshift-corrected
optical and NIR spectrum is shown in Fig. \ref{fig3}. Gaps in the NIR
spectrum are regions of strong absorption by the telluric lines.

Emission-line fluxes in both the optical and NIR ranges were measured by using 
the SPLOT routine in IRAF. 
The errors of the line fluxes are based on the photon statistics
in the non-flux-calibrated spectra. Additionally, we have also included
 an error of 10\% to all lines. This error is that of the 
[S {\sc iii}] 0.907 $\mu$m emission line flux, and including 
it takes into account the fact that this line was used
to adjust the NIR spectrum to the optical one.  
The observed fluxes $F$($\lambda$) of emission lines derived from the adjusted 
optical and NIR spectra 
are shown in the second columns of Tables \ref{tab1} and \ref{tab2}. 
They are normalized to the observed H$\beta$ flux, $F$(H$\beta$) = 
(2.305$\pm$0.008)$\times$10$^{-13}$ erg s$^{-1}$ cm$^{-2}$, and multiplied 
by a factor of 100. In Table
\ref{tab2}, we have indicated the emission lines that are affected by telluric 
absorption. 
Note that several important emission lines in the
optical SDSS spectrum are absent: [O {\sc ii}] 0.373 $\mu$m is out of the
wavelength range of the SDSS spectrum, while the H$\alpha$ 0.656 $\mu$m
and [O {\sc iii}] 0.501 $\mu$m emission lines are clipped. For these
lines, we have adopted the observed fluxes (relative to H$\beta$) derived by 
\citet{I97}. This is a valid procedure because the relative intensities of 
the common emission lines in the SDSS and \citet{I97} spectra are in good 
agreement within the errors. 

Previously, Mrk 59 has been observed in the NIR by
\citet{VR97}. The line fluxes 
of the three emission lines they detected (Br$\gamma$, He {\sc i} 2.06 $\mu$m 
and H$_2$ 2.12 $\mu$m) are lower than ours by a factor of 2.5 -- 3.0. However,
the continuum level in the \citet{VR97} spectrum is a factor of
two higher than that in our spectrum. While the higher continuum level can be 
explained by the larger aperture used by \citet{VR97} 
(2\farcs4$\times$12\arcsec\ as compared to our 1\farcs1$\times$6\arcsec\ 
extraction aperture), aperture effects 
cannot account for the lower line fluxes. 
The reasons for such differences are not clear.
\citet{C96} \citep[see also ][]{C97} have also observed the Br$\gamma$ and 
[Fe {\sc ii}] 1.257 $\mu$m lines in Mrk 59. Their considerably larger aperture 
(6\farcs9$\times$20\arcsec) gives continuum and emission line fluxes that are 
4-5 times larger than our fluxes. However, 
the Br$\gamma$ to [Fe {\sc ii}] 1.257 $\mu$m line ratio is consistent with 
ours within the errors. We obtain  
$F$(Br$\gamma$)/$F$([Fe {\sc ii}]1.257) = 10$\pm$5 
while \citet{C97} gets 15$\pm$6.  

The presence of many emission lines in the optical and NIR spectra
(Figs. \ref{fig2}, \ref{fig3} and Tables \ref{tab1} and \ref{tab2}) 
allows us to study  
extinction and star formation, H$_2$ and Fe {\sc ii} emission  
and excitation mechanisms in Mrk 59. We discuss these issues in the 
following sections.

\subsection{Extinction and star formation}

The fact that the NIR spectrum of Mrk 59 has been obtained simultaneously 
over the entire 
$JHK$ wavelength  
range and that it is possible to directly match the optical spectrum to it, 
using common emission line fluxes in both spectra, eliminates uncertainties introduced 
by the use of different apertures in the optical and NIR observations and
by the adjusting of the continuum levels of 
NIR spectra obtained separately in different orders.   
 In all previous studies of BCDs, the NIR spectra
of BCDs were obtained in separate $JHK$
observations \citep[e.g., ][]{V02}, and there was no wavelength overlap between the 
optical and NIR spectra. The elimination of these adjusting uncertainties permits us to 
compare directly the optical and NIR extinctions.

In the second column of Table \ref{tab3} we show the observed fluxes $F$($\lambda$) 
of the strongest hydrogen lines in both optical and NIR spectra which are
not affected by telluric absorption. 
For comparison, the fourth column of Table \ref{tab3} shows theoretical 
recombination flux ratios, calculated by \citet{HS87} for case B with
an electron temperature $T_e$ = 15000 K and an electron number density
$N_e$ = 100 cm$^{-3}$, which approximate well the values observed 
in Mrk 59 \citep[e.g., ][]{I97}. It is seen from 
Table \ref{tab2} that the  
observed and theoretical ratios are close to
each other for all hydrogen lines, suggesting very little extinction. 
Adopting the \citet{C89} reddening curve with $R(V)$ = $A(V)$/$E(B-V)$ = 3.2,  
we derive an extinction $C$(H$\beta$) = 0.11 from the optical spectrum 
(Table \ref{tab1}). This value of $C$(H$\beta$) corresponds 
to $A(V)$ = 0.24 mag, consistent
with the value $A(V)$ = 0.2 mag obtained by \citet{I97}. The extinction-corrected 
fluxes $I$($\lambda$) of the strong hydrogen lines are shown in the third
column of Table \ref{tab3}. There is a very good agreement between
the corrected and theoretical recombination values of the line fluxes in both
the optical and NIR ranges. This implies that a single $C$(H$\beta$) 
[or $A(V)$] can be used over the whole 0.38 -- 2.40 $\mu$m range to correct line fluxes for extinction. 
The extinction-corrected fluxes $I$($\lambda$) of all emission lines are shown 
in the third column of Tables \ref{tab1} and \ref{tab2} respectively
for the optical and NIR ranges.
The fact that $A(V)$ does not increase when going from the optical to the 
NIR wavelength ranges implies that the NIR light does not probe more extinct regions with 
hidden star formation as compared to the optical light. 
This appears to be a general result for BCDs 
\citep[e.g.][]{V00,V02}. 
Even in the extreme case of the BCD SBS 0335--052E 
 where far-infrared observations imply a very large extinction 
[$A(V)$ $\sim$ 15--20 mag] and three times as much hidden star formation 
as seen in the visible \citep{T99,H04}, 
the $A(V)$ = 0.73 mag derived in the NIR is only slightly greater than the 
optical $A(V)$ = 0.55 mag \citep{V00}.

Using the equivalent width of 154 \AA\ of the H$\beta$ line, Starburst99 
\citep{L99} 
gives a cluster age of about 4.5 Myr in the instantaneous burst case, in good 
agreement with the age  
derived by \citet{B94} for Mrk 59. The presence of Wolf-Rayet stars in Mrk 59 
\citep{G00} also implies that the age of the youngest stellar clusters 
is greater than 3 Myr.  
Another age indicator 
is the clear presence of the CO$\lambda$2.3$\mu$m absorption
feature. Models \citep{L99} 
predict the appearance of this feature only after 10 Myr. This 
implies the presence of slightly older clusters in Mrk 59 which do not 
contribute substantially to the ionization of the gas.
 
The total H$\alpha$ luminosity of Mrk 59 corrected for both Galactic and 
internal extinction at a distance of 10.7 Mpc is  
7.4 $\times$10$^{39}$
ergs s$^{-1}$ \citep{G03}. This corresponds to a star formation rate (SFR) of 
stars more massive than 5 $M_{\odot}$ of 
0.017 $M_{\odot}$ yr$^{-1}$ \citep{C92}. 
The total SFR including all stars more massive than 0.1 $M_{\odot}$ is 
0.09 $M_{\odot}$ yr$^{-1}$, typical of SFRs of BCDs. 
 The thermal radio continuum flux at 1.4 GHz corresponding 
to the extinction-corrected H$\beta$ flux is expected to be about 6 mJy 
\citep{C92}.
\citet{T02a} have observed a 1.4 GHz flux of 10 mJy in Mrk 59.  
Thus most of the radio continuum emission is of thermal nature. The remaining 
flux probably comes from supernova remnants or some hidden star formation.
We favor the latter hypothesis as the spectral index is small, 
equal to $-$0.15 \citep{T02a}, while that of supernova remnants is larger, 
$\sim$ $-$0.8.

\subsection{H$_2$ emission}

Molecular hydrogen lines do not originate in the H {\sc ii} region, but in 
neutral molecular clouds. In the near-infrared, they are excited through two
mechanisms. The first one is the thermal mechanism consisting of 
collisions between neutral species (e.g., H, H$_2$) resulting from large-scale shocks 
driven by powerful stellar winds, supernova remnants 
or molecular cloud collisions \citep[e.g. ][]{SB82}. The second one is 
the fluorescent mechanism due to the absorption of ultraviolet photons
\citep[e.g. ][]{BD76}.
It is known that these two mechanisms mostly excite different
roto-vibrational levels of H$_2$. By comparing the observed line ratios with 
those predicted by models, such as those calculated by \citet{BD87}, it is 
possible to discriminate between the two processes. 
In particular, line emission from the 
vibrational level $v$=2 is virtually absent in collisionally 
excited spectra, while they are relatively strong in fluorescent
spectra.

We detect four H$_2$ emission lines in our NIR spectrum: the 2.034 $\mu$m 1-0 S(2),
the 2.122 $\mu$m 1-0 S(1), the 2.223 $\mu$m 1-0 S(0) and the 2.248 $\mu$m 2-1 S(1) lines. 
The 2.034/2.122, 2.223/2.122 and 2.248/2.122 line flux ratios are close to 0.5
(Table \ref{tab2}), in agreement with the values expected for fluorescent excitation, but  
significantly higher than those predicted for thermal 
excitation \citep{BD87}. In particular, the presence of the 2.248 $\mu$m 2-1 S(1)
line strongly favors the fluorescent mechanism.
We conclude that the H$_2$ lines in Mrk 59 are mainly excited by fluorescence.
\citet{V08} also found the fluorescence excitation process to be  
responsible for the presence of H$_2$ lines in II Zw 40. This galaxy has an 
equivalent width EW(H$\beta$) = 258\AA\ \citep{G06}, implying that its 
starburst has a slightly younger age of $\sim$ 3 Myr. 
How do we reconcile the fact that H$_2$ emission lines are seen in the NIR 
spectrum while \citet{T02b} did not detect any  H$_2$ absorption 
line in their {\sl FUSE} spectrum of Mrk 59? The situation is similar to the 
one in the BCD SBS 0335--052E \citep{T05} where H$_2$ emission is detected in 
the 
NIR, but no UV absorption lines are seen. The explanation for the lack 
of UV absorption lines in Mrk 59 is the same as for SBS 0335--052E: 
the NIR  H$_2$ emission does not come from a uniform low-density 
neutral interstellar medium but from dense clumps. {\sl FUSE} observations 
are not sensitive to such a clumpy  H$_2$ distribution. They can only probe 
diffuse H$_2$ along the lines of sight to the several hundreds of 
UV-bright massive stars in Mrk 59. As in the case of 
other BCDs, the absence of diffuse H$_2$ in Mrk 59 can be explained by the 
combined effects of a low H {\sc i} density, a scarcity of dust grains on 
which H$_2$ molecules can form, and a large UV flux that destroys molecules
\citep{T02b}.

\subsection{Stellar photoionization CLOUDY model of the H {\sc ii} region}

We next examine the excitation mechanisms for the NIR emission lines in
the H {\sc ii} region of Mrk 59. For this purpose, we have constructed 
a stellar photoionization model for the H {\sc ii} region, using
the CLOUDY code of \citet{F96} and \citet{F98}.  
We list in Table \ref{tab4} the input parameters of this model. 
$Q$(H) is the log of
the number of ionizing photons per second, 
calculated from the H$\beta$ luminosity
at a distance of 10.7 Mpc \citep{T02a}, 
$T_{eff}$ is the effective temperature
of stellar radiation, $N_e$ is the electron number density and $f$ is 
the filling factor. The remaining input parameters are the number ratios of
different species to hydrogen. We varied the abundances
of different elements to obtain agreement between the observed
and predicted line fluxes. Exceptions are the oxygen abundance
for which we use the value derived from the SDSS spectrum, and the carbon
and silicon abundances. There are no strong emission lines
of these species in the optical and NIR ranges. We have therefore taken 
their abundances from \citet{KS98}.
For the ionizing stellar radiation, 
we have adopted a \citet{K79} stellar atmosphere model with $T_{eff}$=50000 K. 
We have also experimented with other forms of the ionizing stellar
radiation such as those given by Starburst99 models \citep{L99} or by 
models based on Geneva stellar evolutionary tracks with a heavy element
mass fraction $Z$ = 0.004 \citep{M94}. 
In all cases, we were able to reproduce well the observed line 
intensities by varying only slightly the input parameters. Thus 
our results are robust and do not depend sensitively on a particular  
adopted grid of models.

The CLOUDY-predicted
fluxes of the optical 
emission lines are shown in the fourth column of Table \ref{tab1}.
Comparison of the third and fourth columns of that table 
shows that, in general,
the agreement between the optical 
observations and the CLOUDY predictions is very good.
The most notable difference is for sulfur. The model reproduces well 
the [S {\sc iii}] emission line fluxes, but the predicted fluxes
for the [S {\sc ii}] nebular lines are lower by a factor of $\sim$ 2 
than the observed ones. A probable cause for this discrepancy is the poorly 
known dielectronic recombination coefficients of sulfur, on 
which the ionic abundances depend sensitively 
\citep[e.g., ][]{I06}.

Examination of the third and fourth columns of Table \ref{tab2} shows that 
the same CLOUDY model reproduces also very well the observed emission-line 
fluxes in the NIR spectrum.
This agreement implies that a H {\sc ii} region model including only stellar 
photoionization as ionizing source is able to 
account for the observed fluxes both in the optical and NIR ranges.
No additional excitation mechanisms such as shocks from 
stellar winds and supernova remnants 
are needed. This situation is again similar to that  in the BCD II Zw 40 where 
\citet{V08} found that the interstellar medium (ISM) is mainly 
photoexcited by stars.

\subsection{[Fe {\sc ii}] line emission}

The [Fe {\sc ii}] 1.257 and 1.643 $\mu$m emission
lines are clearly detected (Fig. \ref{fig1}). They have  
often been used as indicators of the importance of shock excitation 
relative to photoionization \citep{MO88,O90,O01}. 
In the Galaxy, 
the intensities of these lines are considerably 
smaller in regions where photoionization dominates than in those 
where shock excitation dominates, such as in SN remnants. For example,      
the ratio of the [Fe {\sc ii}] 1.643 $\mu$m line to Br$\gamma$ is 0.06 in the 
Orion nebula, while it is more than 30 in Galactic SN remnants 
\citep{MO88,O90}. This is because fast shocks propagating in the 
diffuse ISM can destroy dust grains through sputtering processes or 
grain-grain collisions,and replenish the ISM with gas-phase iron. 
The gaseous iron is then collisionally excited in the cooling post-shock gas, 
and produces the observed infrared emission.    

In the Mrk 59 spectrum, the ratios of the 
[Fe {\sc ii}] 1.257 $\mu$m and [Fe {\sc ii}] 1.643 $\mu$m line intensities to 
that of Br$\gamma$ have the very low 
values of 0.099 and 0.118, respectively.
\citet{C97} also found that the [Fe {\sc ii}] 1.257 $\mu$m line flux is 
some 30 times smaller in Mrk 59 (for a given Pa$\beta$ flux) as compared to 
the remaining galaxies in her sample of 19 starburst galaxies, all 
having higher metallicity (about solar) and lower excitation.   
Does this mean that there is no significant
SN shock excitation in Mrk 59?  
If this is the case, this raises a puzzle since the age of 
the ionizing cluster in Mrk 59 is 4.5 Myr. For a Salpeter IMF and 
an upper mass limit of 100 $M_\odot$, 
Starburst99 \citep{L99} gives a SN rate of 1.4$\times$10$^{-4}$ yr$^{-1}$.
Since the first SN 
explodes around 3 Myr, we would expect $\sim$ 200 SNe to have occurred 
by the age of 4.5 Myr. 

Examination of Table \ref{tab2} shows that the stellar photoionization 
CLOUDY model reproduces reasonably well the observed [Fe {\sc ii}] 1.257 and 
1.643 $\mu$m line fluxes. We have also run a CLOUDY model where 
the ionization radiation does not come from stars like in the 
model discussed in Section 3.3, but from fast radiative shocks, as modeled
by \citet{A08}. 
We use the slowest available shock model from \citet{A08},
with a velocity of 100 km s$^{-1}$. With the observed 
log Q = 51.953 (Table \ref{tab4}),  
we again obtain [Fe {\sc ii}] 1.257 $\mu$m/Br$\gamma$ and 
1.643 $\mu$m/Br$\gamma$ line flux ratios 
of $\sim$ 0.1, similar to those obtained with the model with 
stellar ionizing radiation (Table \ref{tab2}). It is clear that, 
in a high-excitation H {\sc ii} 
region as Mrk 59, the 
[Fe {\sc ii}] 1.257 and 1.643 $\mu$m emission lines are not particularly 
sensitive to the presence of shocks, since CLOUDY models with a stellar 
ionizing spectrum or with a shock ionizing spectrum produce the same line 
intensities.   
However, the shock model fails to reproduce the 
strong forbidden emission line fluxes, for example 
overpredicting the  
[O {\sc ii}] 0.373 $\mu$m emission line by 40\% and underpredicting  
the [O {\sc iii}] 0.501 $\mu$m emisssion line by 80\%. 
It is clear that shocks, although likely present in Mrk 59,
 cannot be the main source of ionization in it. Stellar 
radiation is.   

We thus conclude that the [Fe {\sc ii}] 1.257 and 1.643 $\mu$m emission lines 
cannot  
be used as shock indicators in high-excitation low-metallicity 
H {\sc ii} regions. This means, for example, that 
the Starburst99 \citep{L99} predictions for the [Fe {\sc ii}] 1.257
line flux as a function of the supernova rate cannot be applied to Mrk
59. This is because these
predictions are based on mean relations established by \citet{C97} for 
[Fe {\sc ii}] line emission in low-excitation high-metallicity starburst 
galaxies. \citet{C97} found that the [Fe {\sc ii}] 1.257
line flux of the high-excitation low-metallicity BCD 
Mrk 59 (= NGC 4861) deviates strongly from that mean relation, 
being a factor of 30 below the [Fe {\sc ii}] 1.257
line flux -- Pa$\beta$ line flux mean relation.      
This could be due to two reasons: 1) the lower metallicity in Mrk 59 
results in a lower Fe abundance; and 2) the higher excitation of Mrk 59 
results in an iron ionic composition which is very different from that for 
other galaxies. 
Our CLOUDY stellar photoionization model 
predicts that most of the iron in Mrk 59 
is in the Fe {\sc iv} 
ionization stage, there being about 10 times more Fe {\sc iv} ions 
than Fe {\sc ii} ions (by volume), and about 3 times more  Fe {\sc iii} ions 
than Fe {\sc ii} ions.

\section{CONCLUSIONS}

We discuss here the results of near-infrared (NIR) spectroscopic observations
of the blue compact dwarf (BCD) galaxy Mrk 59. The NIR data have been 
supplemented
by the optical spectrum 
from the Sloan Digital Sky Survey (SDSS), resulting
in a total wavelength coverage of 0.38 -- 2.46 $\mu$m. The 
overlap in the 0.90 - 0.93 $\mu$m wavelength range allows to adjust 
the flux level of the NIR spectrum to that of the optical spectrum, 
using fluxes of common emission lines.

We have arrived at the following conclusions:

1. Using the Balmer decrement of hydrogen emission lines in the optical
range, we find that the extinction in Mrk 59 is low, amounting to $A(V)$ =
0.24 mag. Correcting the observed 
hydrogen emission line fluxes in the NIR range 
for this optically-derived reddening gives values that are 
very close to the theoretical
recombination values. Thus, extinction in Mrk 59 is the 
same in both optical and NIR ranges. 
The NIR light does not probe more extinct regions with 
hidden star formation as compared to the optical light.  

2. We have detected
 four molecular hydrogen emission lines in the NIR spectrum of 
Mrk 59. Comparison 
of the observed fluxes with modeled ones 
suggests that the main excitation mechanism of H$_2$ emission
in Mrk 59 is fluorescence. The H$_2$ emission comes not from a 
uniform low-density neutral medium but from dense clumps. 

3. A CLOUDY model with a pure stellar ionizing radiation
reproduces well the observed emission line fluxes in both optical and NIR
ranges. Shocks are likely present in the H {\sc ii} region, but 
play a minor role in the ionization. 

4. CLOUDY stellar ionizing and shock excitation models show that   
the [Fe {\sc ii}] 1.257 and 1.643 $\mu$m emission lines, often used 
as SN shock indicators in low-excitation high-metallicity starburst galaxies, 
cannot play such a role in high-excitation low-metallicity 
H {\sc ii} regions such as Mrk 59. Both types of 
models predict similarly low [Fe {\sc ii}] emission line fluxes for 
the latter category of objects, 
so that these line fluxes cannot discriminate between the two mechanisms.   

\acknowledgements
We thank The TripleSpec team for their superb
work in building TripleSpec and the APO staff for their 
valuable assistance during commissioning. 
Y.I.I. is grateful to the staff of the Astronomy Department at the 
University of Virginia for their warm hospitality. 
Support for this work is provided by 
NASA through contract 1263707 issued by JPL/Caltech.

\clearpage

\clearpage

  \begin{deluxetable}{lrrr}
  \tablecolumns{2}
  \tablewidth{0pc}
  \tablecaption{Fluxes of optical emission lines
\label{tab1}}
  \tablehead{
  \colhead{\sc Ion}
  &\colhead{100$\times$$F$($\lambda$)/$F$(H$\beta$)\tablenotemark{a}}
&\colhead{100$\times$$I$($\lambda$)/$I$(H$\beta$)\tablenotemark{b}}
&\colhead{CLOUDY}
}
  \startdata
0.373 [O {\sc  ii}]\tablenotemark{c}&  104.70$\pm$3.00 &  113.24$\pm$3.50&124.16 \\
0.384 H9                            &    6.31$\pm$0.23 &    6.90$\pm$0.29&  7.69 \\
0.387 [Ne {\sc  iii}]               &   45.49$\pm$1.32 &   48.83$\pm$1.51& 49.80 \\
0.389 He {\sc i}+H8                 &   18.14$\pm$0.55 &   19.56$\pm$0.64& 19.86 \\
0.389 [Ne {\sc iii}]+H7             &   28.12$\pm$0.83 &   30.12$\pm$0.94& 31.31 \\
0.403 He {\sc i}                    &    1.35$\pm$0.12 &    1.43$\pm$0.13&  1.91 \\
0.407 [S {\sc ii}]                  &    0.92$\pm$0.12 &    0.97$\pm$0.12&  0.63 \\
0.408 [S {\sc ii}]                  &    0.20$\pm$0.11 &    0.21$\pm$0.12&  0.21 \\
0.384 H$\delta$                     &   24.72$\pm$0.73 &   26.24$\pm$0.81& 26.31 \\
0.434 H$\gamma$                     &   46.83$\pm$1.36 &   48.73$\pm$1.44& 47.37 \\
0.436 [O {\sc iii}]                 &    8.53$\pm$0.28 &    8.84$\pm$0.29&  8.85 \\
0.439 He {\sc i}                    &    0.35$\pm$0.10 &    0.36$\pm$0.10&  0.51 \\
0.447 He {\sc i}                    &    3.48$\pm$0.15 &    3.58$\pm$0.15&  3.95 \\
0.466 [Fe {\sc  iii}]               &    0.60$\pm$0.10 &    0.61$\pm$0.10&  0.64 \\
0.471 [Ar {\sc iv}]+He {\sc i}      &    1.26$\pm$0.10 &    1.28$\pm$0.10&  1.28 \\
0.474 [Ar {\sc iv}]                 &    0.71$\pm$0.10 &    0.71$\pm$0.10&  0.59 \\
0.486 H$\beta$                      &  100.00$\pm$2.87 &  100.00$\pm$2.87&100.00 \\
0.492 He {\sc i}                    &    0.80$\pm$0.09 &    0.80$\pm$0.09&  1.06 \\
0.496 [O {\sc iii}]                 &  207.14$\pm$5.92 &  205.70$\pm$5.89&197.09 \\
0.499 [Fe {\sc  iii}]               &    0.58$\pm$0.09 &    0.58$\pm$0.09&  0.02 \\
0.501 [O {\sc iii}]\tablenotemark{c}&  594.00$\pm$17.0 &  588.49$\pm$16.9&593.23 \\
0.502 He {\sc i}                    &    2.27$\pm$0.12 &    2.24$\pm$0.12&  2.53 \\
0.520 [N {\sc  i}]                  &    0.33$\pm$0.08 &    0.32$\pm$0.08&  0.09 \\
0.527 [Fe {\sc  iii}]               &    0.26$\pm$0.08 &    0.26$\pm$0.08&  0.32 \\
0.552 [Cl {\sc  iii}]               &    0.38$\pm$0.08 &    0.36$\pm$0.07&  0.36 \\
0.554 [Cl {\sc  iii}]               &    0.29$\pm$0.08 &    0.28$\pm$0.07&  0.27 \\
0.575 [N {\sc  ii}]                 &    0.13$\pm$0.07 &    0.13$\pm$0.07&  0.11 \\
0.588 He {\sc i}                    &   10.52$\pm$0.33 &    9.99$\pm$0.32&  9.91 \\
0.630 [O {\sc  i}]                  &    1.77$\pm$0.09 &    1.66$\pm$0.09&  1.57 \\
0.631 [S {\sc iii}]                 &    1.96$\pm$0.09 &    1.84$\pm$0.09&  1.76 \\
0.636 [O {\sc  i}]                  &    0.56$\pm$0.07 &    0.52$\pm$0.06&  0.50 \\
0.655 [N {\sc  ii}]                 &    1.70$\pm$0.09 &    1.58$\pm$0.08&  1.58 \\
0.656 H$\alpha$\tablenotemark{c}    &  307.20$\pm$8.77 &  285.15$\pm$8.71&283.73 \\
0.658 [N {\sc  ii}]                 &    4.68$\pm$0.16 &    4.34$\pm$0.16&  4.65 \\
0.668 He {\sc i}                    &    3.06$\pm$0.12 &    2.83$\pm$0.12&  2.74 \\
0.672 [S {\sc  ii}]                 &    9.85$\pm$0.30 &    9.09$\pm$0.30&  4.74 \\
0.673 [S {\sc  ii}]                 &    7.22$\pm$0.23 &    6.67$\pm$0.23&  3.59 \\
0.707 He {\sc i}                    &    2.52$\pm$0.10 &    2.30$\pm$0.10&  3.15 \\
0.714 [Ar {\sc  iii}]               &    8.65$\pm$0.27 &    7.87$\pm$0.27&  7.83 \\
0.728 He {\sc i}                    &    0.64$\pm$0.06 &    0.58$\pm$0.06&  0.67 \\
0.732 [O {\sc  ii}]                 &    1.59$\pm$0.08 &    1.44$\pm$0.08&  2.18 \\
0.733 [O {\sc  ii}]                 &    1.31$\pm$0.07 &    1.18$\pm$0.07&  1.73 \\
0.775 [Ar {\sc  iii}]               &    2.02$\pm$0.09 &    1.80$\pm$0.09&  1.89 \\
0.847 Pa17                           &    0.37$\pm$0.15 &    0.32$\pm$0.14&  0.29 \\
0.850 Pa16                           &    0.47$\pm$0.15 &    0.41$\pm$0.14&  0.34 \\
0.855 Pa15                           &    0.47$\pm$0.15 &    0.41$\pm$0.14&  0.45 \\
0.860 Pa14                           &    0.71$\pm$0.15 &    0.62$\pm$0.14&  0.99 \\
0.867 Pa13                           &    0.73$\pm$0.15 &    0.63$\pm$0.14&  1.03 \\
0.875 Pa12                           &    1.08$\pm$0.15 &    0.95$\pm$0.14&  1.20 \\
0.886 Pa11                           &    1.46$\pm$0.17 &    1.26$\pm$0.16&  1.49 \\
0.901 Pa10                           &    2.08$\pm$0.19 &    1.79$\pm$0.18&  1.92 \\
0.907 [S {\sc iii}]                 &   16.96$\pm$0.50 &   14.62$\pm$0.54& 15.06 \\
  \enddata
\tablenotetext{a}{$F$(H$\beta$)=(2.305$\pm$0.008)$\times$10$^{-13}$ erg s$^{-1}$ cm$^{-2}$.}
\tablenotetext{b}{Corrected for extinction with $C$(H$\beta$)=0.11.}
\tablenotetext{c}{Observed flux is from \citet{I97}.}
  \end{deluxetable}

\clearpage

  \begin{deluxetable}{lrrr}
  \tablecolumns{4}
  \tablewidth{0pc}
  \tablecaption{Fluxes of near-infrared emission lines
\label{tab2}}
  \tablehead{
  \colhead{\sc Ion}
  &\colhead{100$\times$$F$($\lambda$)/$F$(H$\beta$)\tablenotemark{a}}
&\colhead{100$\times$$I$($\lambda$)/$I$(H$\beta$)\tablenotemark{b}}
&\colhead{CLOUDY}
}
  \startdata
0.907 [S {\sc iii}]                  &   16.98$\pm$1.63 &   14.63$\pm$1.40& 15.06 \\
0.953 [S {\sc iii}]\tablenotemark{c} &   43.43$\pm$4.27 &   37.13$\pm$3.88& 37.35 \\
0.955 Pa$\epsilon$\tablenotemark{c}   &    3.26$\pm$0.43 &    2.79$\pm$0.41&  3.66 \\
1.005 Pa$\delta$                      &    5.44$\pm$0.56 &    4.61$\pm$0.45&  5.47 \\
1.029 [S {\sc ii}]                   &    0.15$\pm$0.06 &    0.13$\pm$0.06&  0.15 \\
1.032 [S {\sc ii}]                   &    0.15$\pm$0.08 &    0.13$\pm$0.08&  0.20 \\
1.083 He {\sc i}                     &   24.44$\pm$2.42 &   20.28$\pm$2.28& 25.96 \\
1.091 He {\sc i}                     &    0.43$\pm$0.08 &    0.34$\pm$0.08&  0.28 \\
1.094 Pa$\gamma$                     &    9.08$\pm$0.91 &    7.61$\pm$0.66&  8.77 \\
1.163 He {\sc ii}                    &    0.12$\pm$0.07 &    0.09$\pm$0.06&  0.00 \\
1.197 He {\sc i}                     &    0.16$\pm$0.07 &    0.14$\pm$0.07&  0.19 \\
1.253 He {\sc i}                     &    0.16$\pm$0.06 &    0.14$\pm$0.06&  0.19 \\
1.257 [Fe {\sc ii}]                  &    0.28$\pm$0.09 &    0.25$\pm$0.09&  0.34 \\
1.279 He {\sc i}                     &    0.96$\pm$0.12 &    0.94$\pm$0.11&  0.53 \\
1.282 Pa$\beta$                      &   16.40$\pm$1.60 &   14.64$\pm$1.45& 15.65 \\
1.544 Br17                           &    0.22$\pm$0.07 &    0.18$\pm$0.06&  0.14 \\
1.556 Br16                           &    0.24$\pm$0.07 &    0.19$\pm$0.06&  0.17 \\
1.570 Br15                           &    0.30$\pm$0.09 &    0.24$\pm$0.07&  0.22 \\
1.588 Br14                           &    0.35$\pm$0.07 &    0.29$\pm$0.07&  0.47 \\
1.611 Br13                           &    0.45$\pm$0.08 &    0.36$\pm$0.07&  0.49 \\
1.641 Br12                           &    0.53$\pm$0.08 &    0.43$\pm$0.08&  0.58 \\
1.643 [Fe {\sc ii}]                  &    0.36$\pm$0.07 &    0.30$\pm$0.07&  0.33 \\
1.681 Br11                           &    0.78$\pm$0.11 &    0.63$\pm$0.10&  0.71 \\
1.700 He {\sc i}                     &    0.38$\pm$0.09 &    0.30$\pm$0.07&  0.28 \\
1.732 He {\sc i}                     &    0.10$\pm$0.04 &    0.08$\pm$0.03&\nodata\\
1.736 Br10                           &    1.02$\pm$0.13 &    0.82$\pm$0.09&  0.92 \\
1.745 [Fe {\sc ii}]+He {\sc i}       &    0.07$\pm$0.05 &    0.06$\pm$0.04&\nodata\\
1.876 Pa$\alpha$\tablenotemark{c}     &  33.63$\pm$3.18 &   26.97$\pm$2.78& 31.74 \\
1.944 Br$\delta$\tablenotemark{c}    &    2.16$\pm$0.37 &    1.73$\pm$0.28&  1.77 \\
2.034 H$_2$ 1-0 S(2)                 &    0.13$\pm$0.04 &    0.10$\pm$0.04&\nodata\\
2.058 He {\sc i}\tablenotemark{c}    &    1.32$\pm$0.16 &    1.07$\pm$0.14&  1.45 \\
2.122 H$_2$ 1-0 S(1)                 &    0.26$\pm$0.09 &    0.21$\pm$0.08&\nodata\\
2.165 Br$\gamma$                     &    3.18$\pm$0.32 &    2.54$\pm$0.30&  2.64 \\
2.223 H$_2$ 1-0 S(0)                 &    0.10$\pm$0.06 &    0.08$\pm$0.05&\nodata\\
2.248 H$_2$ 2-1 S(1)                 &    0.15$\pm$0.06 &    0.11$\pm$0.05&\nodata\\
  \enddata
\tablenotetext{a}{$F$(H$\beta$)=(2.305$\pm$0.008)$\times$10$^{-13}$ erg s$^{-1}$ cm$^{-2}$.}
\tablenotetext{b}{Corrected for extinction with $C$(H$\beta$)=0.11.}
\tablenotetext{c}{The observed flux is affected by telluric absorption.}
  \end{deluxetable}

\clearpage

\begin{deluxetable}{lrrr}
  \tablecolumns{4}
  \tablewidth{0pt}
  \tablecaption{Fluxes of  strong hydrogen lines \label{tab3}}
  \tablehead{
\colhead{Line}
&\multicolumn{1}{c}{100$\times$$F$($\lambda$)/$F$(H$\beta$)\tablenotemark{a}}
&\multicolumn{1}{c}{100$\times$$I$($\lambda$)/$I$(H$\beta$)\tablenotemark{b}}
&\colhead{Theor.\tablenotemark{c}}
}
  \startdata
\multicolumn{4}{c}{a) Optical spectrum} \\
0.410 H$\delta$      & 24.7 &  26.2 & 26.2 \\
0.434 H$\gamma$      & 46.8 &  48.7 & 47.3 \\
0.434 H$\beta$       &100.0 & 100.0 &100.0 \\
0.875 Pa12            &  1.1 &   1.0 &  1.0 \\
0.886 Pa11            &  1.5 &   1.3 &  1.3 \\
0.901 Pa10            &  2.1 &   1.8 &  1.8 \\
\multicolumn{4}{c}{b) NIR spectrum} \\
1.094 Pa$\gamma$      &  9.1 &   7.6 &  8.6 \\
1.282 Pa$\beta$       & 16.4 &  14.6 & 15.2 \\
2.165 Br$\gamma$     &  3.2 &   2.5 &  2.5 \\
  \enddata
\tablenotetext{a}{$F$(H$\beta$)=2.305$\times$10$^{-13}$ 
ergs s$^{-1}$cm$^{-2}$.}
\tablenotetext{b}{Corrected for extinction with $C$(H$\beta$)=0.11.}
\tablenotetext{c}{Recombination ratios for $T_e$=15000 K and 
$N_e$ = 100 cm$^{-3}$ from \citet{HS87}.}
  \end{deluxetable}

\clearpage

\begin{deluxetable}{lr}
  \tablecolumns{4}
  \tablewidth{0pt}
  \tablecaption{Input parameters for stellar photoionization 
CLOUDY model \label{tab4}}
  \tablehead{
\colhead{Parameter}&\colhead{Value}
}
  \startdata
$Q$(H)                 & 51.953 \\
$T_{eff}$, K           & 50000  \\
$N_e$, cm$^{-3}$       & 100    \\
$f$                    & 0.07   \\
log He/H               & --1.07 \\
log C/H                & --4.48 \\
log N/H                & --5.34 \\
log O/H                & --3.97 \\
log Ne/H               & --4.72 \\
log Si/H               & --5.35 \\
log S/H                & --5.69 \\
log Cl/H               & --7.59 \\
log Ar/H               & --6.32 \\
log Fe/H               & --6.22 \\
  \enddata
  \end{deluxetable}

\clearpage

\clearpage

\clearpage

\begin{thebibliography}{}

\bibitem[Allen et al. (2008)]{A08} Allen, M. G., Groves, B. A., Dopita, M. A.,
Sutherland, R. S., \& Kewley, L. J. 2008, \apjs, 178, 20


\bibitem[Arp (1966)]{A66} Arp, H. C. 1966, \apjs, 14, 1

\bibitem[Barth et al. (1994)]{B94} Barth, C. S., Cepa, J., Vilchez, J. M., 
\& Dottori, H. A. 1994, \aj, 108, 2069

\bibitem[Black \& Dalgarno (1976)]{BD76} Black, J. H., \& Dalgarno, A. 1976,
\apj, 203, 132

\bibitem[Black \& van Dishoeck (1987)]{BD87} Black, J. H., \& van Dishoeck,
E. F. 1987, \apj, 322, 412

\bibitem[Calzetti (1997)]{C97} Calzetti, D. 1997, \aj, 113, 162

\bibitem[Calzetti et al. (1996)]{C96} Calzetti, D., Kinney, A. L., \& 
Storchi-Bergmann, T. 1996, \apj, 458, 132

\bibitem[Cardelli et al. (1989)]{C89} Cardelli, J. A., Clayton, G. C., \&
Mathis, J. S. 1989, \apj, 345, 245

\bibitem[Condon (1992)]{C92} Condon, J. J. 1992, \araa, 30, 575

\bibitem[Dinerstein \& Shields (1986)]{DS86} Dinerstein, H. L., 
\& Shields, G. A. 1986, \apj, 311, 45

\bibitem[Ferland (1996)]{F96} Ferland, G. J. 1996, Hazy: A brief 
Introduction to CLOUDY, Univ. Kentucky Phys. Dept. Int. Rep.

\bibitem[Ferland et al. (1998)]{F98} Ferland, G. J., Korista, K. T., 
Verner, D. A., Ferguson, J. W., Kingdon, J. B., \&
Verner, E. M. 1998, \pasp, 110, 761

\bibitem[Gil de Paz et al. (2003)]{G03} Gil de Paz, A., Madore, B. F., 
\& Pevunova, O. 2003, \apjs, 147, 29

\bibitem[Guseva et al. (2000)]{G00} Guseva, N. G., Izotov, Y. I., 
\& Thuan, T. X. 2000, \apj, 531, 776

\bibitem[Guseva et al. (2006)]{G06} Guseva, N. G., Izotov, Y. I., 
\& Thuan, T. X. 2006, \apj, 644, 890

\bibitem[Houck et al. (2004)]{H04} Houck, J. R. et al. 2004, \apjs, 154, 211

\bibitem[Hummer \& Storey (1987)]{HS87} Hummer, D. G., \& Storey, P. J.
1987, \mnras, 224, 801


\bibitem[Izotov et al. (1997)]{I97} Izotov, Y. I., Thuan, T. X., 
\& Lipovetsky, V. A. 1997, \apjs, 108, 1

\bibitem[Izotov et al. (2006)]{I06} Izotov, Y. I., Stasi\'nska, G., 
Meynet, G., Guseva, N. G., \& Thuan T. X. 2006, \aap, 448, 955

\bibitem[Kobulnicky \& Skillman (1998)]{KS98} Kobulnicky, H. A., \&
Skillman, E. D. 1998, \apj, 497, 601

\bibitem[Kurucz (1979)]{K79} Kurucz, R. L. 1979, \apjs, 40, 1

\bibitem[Leitherer et al. (1999)]{L99} Leitherer, C., et al. 1999, \apjs,
123, 3

\bibitem[Loose \& Thuan (1985)]{LT85} Loose, H.-H., \& Thuan, T. X. 1985, 
in Star-forming Dwarf Galaxies
and Related Objects, ed. D. Kunth, T. X. Thuan \& J. T. T. Van 
(Gif-sur-Yvette: Editions Fronti\`eres), 73

\bibitem[Meynet et al. (1994)]{M94} Meynet, G., Maeder, A., Schaller, G., 
Schaerer, D., \& Charbonnel, C. 1994, \aaps, 103, 97

\bibitem[Moorwood \& Oliva (1988)]{MO88} Moorwood, A. F. M., \& Oliva, E.
1988, \aap, 203, 278

\bibitem[Noeske et al. (2000)]{N00} Noeske, K. G., Guseva, N. G., 
Fricke, K. J., Izotov, Y. I., Papaderos, P., \& Thuan, T. X. 2000, 
\aap, 361, 33

\bibitem[Oliva et al. (1990)]{O90} Oliva, E., Moorwood, A. F. M., \&
Danziger, I. J. 1990, \aap, 240, 453

\bibitem[Oliva et al. (2001)]{O01} Oliva, E., et al. 2001, \aap, 369, L5

\bibitem[Shull \& Beckwith (1982)]{SB82} Shull, J. M., \& Beckwith, S.
1982, \araa, 20, 163

\bibitem[Thuan et al. (1987)]{T87} Thuan, T. X., Williams, T. B., \& 
Malumuth, E. 1987, in Starbursts and Galaxy
Evolution, ed. T. X. Thuan, T. Montmerle, \& J. T. T. Van (Gif-sur-Yvette: 
Editions Fronti\`eres), 151

\bibitem[Thuan et al. (1999)]{T99} Thuan, T. X., Sauvage, M., 
\& Madden, S. 1999, \apj, 516, 783

\bibitem[Thuan et al. (2002a)]{T02a} Thuan, T. X., L\'evrier, F., 
\& Hibbard, J. E. 2002a, \aj, 128, 617

\bibitem[Thuan et al. (2002b)]{T02b} Thuan, T. X., Lecavelier des Etangs, A., 
\& Izotov, Y. I. 2002b, \apj, 565, 941

\bibitem[Thuan et al. (2005)]{T05} Thuan, T. X., Lecavelier des Etangs, A., 
\& Izotov, Y. I. 2005, \apj, 621, 269

\bibitem[Vanzi \& Rieke (1997)]{VR97} Vanzi, L., \& Rieke, G. H. 1997,
\apj, 479, 694


\bibitem[Vanzi et al. (2000)]{V00} Vanzi, L., Hunt, L. K., Thuan, T. X.,
\& Izotov, Y. I. 2000, \aap, 363, 493

\bibitem[Vanzi et al. (2002)]{V02} Vanzi, L., Hunt, L. K., \& Thuan, T. X.
2002, \aap, 390, 481

\bibitem[Vanzi et al. (2008)]{V08} Vanzi, L., Cresci, G., Telles, E., \&
Melnick, J. 2008, \aap, 486, 393

\bibitem[Wilcots et al. (1996)]{W96} Wilcots, E. M., Lehman, C., 
\& Miller, B. 1996, \aj, 111, 1575

\bibitem[Wilson et al. (2004)]{W04} Wilson, J. C., et al. 2004, SPIE, 5492, 
1295 





\end{thebibliography}
\end{document}